\documentclass[prd,showpacs,showkeys]{revtex4}
\usepackage{bm}
\usepackage{graphicx}
\usepackage{amssymb}
\begin{document}
\title{Algorithmic construction of static perfect fluid spheres}
\author{Damien Martin}
\affiliation{School of Mathematical and Computing Sciences, 
Victoria University of Wellington, PO Box 600, Wellington, New Zealand\\}
\author{Matt Visser}
\email{matt.visser@vuw.ac.nz}
\homepage{http://www.mcs.vuw.ac.nz/~visser}
\affiliation{School of Mathematical and Computing Sciences, 
Victoria University of Wellington, PO Box 600, Wellington, New Zealand\\}
\date{24 June 2003; 9 March 2004; 
\LaTeX-ed \today}
\begin{abstract}
  Perfect fluid spheres, either Newtonian and relativistic, are the
  first step in developing realistic stellar models (or models for
  fluid planets). Despite the importance of these models, explicit and
  fully general solutions of the perfect fluid constraint in general
  relativity have only very recently been developed.  In this Brief
  Report we present a variant of Lake's algorithm wherein: (1) we
  re-cast the algorithm in terms of variables with a clear physical
  meaning --- the average density and the locally measured
  acceleration due to gravity, (2) we present explicit and fully
  general formulae for the mass profile and pressure profile, and (3)
  we present an explicit closed-form expression for the central
  pressure. Furthermore we can then use the formalism to easily
  understand the pattern of inter-relationships among many of the
  previously known exact solutions, and generate several new exact
  solutions.


\end{abstract}


\maketitle
\newcommand{\diff}[1]{\ensuremath{\mathrm{d}{#1}}}
\newcommand{\dx}[1]{\diff{#1}}
\newcommand{\dr}{\ensuremath{\frac{\mathrm{d}\phantom{r}}{\dx{r}}}}
\newcommand{\eqprime}[1]{\tag{\ref{#1}$^\prime$}}
\newcommand{\dOne}{\ensuremath{\delta_1}}
\newcommand{\dTwo}{\ensuremath{\delta_2}}
\newcommand{\dThr}{\ensuremath{\delta_3}}
\newtheorem{theorem}{Theorem}
\newtheorem{corollary}{Corollary}
\def\d{{\mathrm{d}}}
\def\implies{\Rightarrow}
\section{Introduction}

Perfect fluid spheres, both Newtonian~\cite{Chandrasekhar} and
relativistic~\cite{Buchdahl,Bondi}, have attracted and continue to
attract considerable attention as the first step in developing
realistic stellar models (or models for fluid planets). Whereas some
steps toward finding all possible solutions to the perfect fluid
constraint in the absence of a specific equation of state were
presented in work early of Wyman and Hojman \emph{et
  al}~\cite{Wyman,Hojman-et-al}, \emph{explicit} and fully general
solutions of the perfect fluid constraint have only very recently been
developed~\cite{Rahman,Lake}. In this article we present a variant of
Lake's algorithm~\cite{Lake} using curvature coordinates wherein:
\begin{itemize}
\item We re-cast the algorithm in terms of variables with a clear
  physical meaning --- the average density and the ``gravity
  profile'', a quantity closely related to both the gravitational
  redshift and the locally measured acceleration due to gravity.
\item We minimize the number of differentiations and integrations by
  several judicious applications of integration by parts.
\item We present explicit, compact, and fully general formulae for the
  mass profile and pressure profile of an \emph{arbitrary} fluid sphere.
\item We present an explicit, compact, and general formula for the
  central pressure of an \emph{arbitrary} fluid sphere.
\item We compare and contrast the relativistic formulae we obtain with
  the much simpler Newtonian situation.
\end{itemize}
We emphasise that one of the virtues of this type of approach is that
one is not fixed \emph{a priori} to dealing with a pre-specified
equation of state~\cite{Rahman} --- in many interesting physical
situations the equation of state is either uncertain or, because the
fluid in question might be inhomogeneous, it may not even make sense to
assign a single equation of state to the entire fluid sphere.

To further illustrate the formalism we show how it may be used as the
basis for a partial classification scheme --- there is a free
parameter in the algorithm that can take simple solutions into more
complicated ones. Once this is appreciated it becomes easy to see
(simply by parameter counting) that certain simple solutions
\emph{must} have one-parameter generalizations. Conversely, this
observation explains why so many of the earliest discovered exact
solutions have one-parameter extensions.

\section{Framework}

To set the stage, consider a static spherically symmetric geometry. It
is a standard result that without loss of generality we can choose
coordinates to write the metric in the form
\begin{equation}
\d s^2 = - \;\exp\left[-2\int_r^\infty g(\tilde r) \; \d\tilde r \right] \; \d t^2
+ {\d r^2 \over 1-2m(r)/r} +
r^2 \left[ \d\theta^2 +\sin^2\theta\;\d\phi^2\right].
\end{equation}
Here $g(r)$ ``gravity profile''. It is related to the gravitational
redshift by 
\begin{equation}
1+z = \exp\left[\int_r^\infty g(\tilde r) \d \tilde r\right],
\end{equation}
and is related to the locally measured acceleration due to gravity by
\begin{equation}
a=\sqrt{1-{2m(r)\over r}} \; g(r).
\end{equation} 
Our convention is that $g(r)$ is positive for a downward
acceleration. The function $m(r)$ is the quasi-local mass. In the
vacuum region beyond the surface (if any) of the star-like object, the
Schwarzschild solution yields $g(r) = {(M/r^2)/(1 - 2M/r)}$ and $m(r)
= M$.  We find it more convenient to write the metric in the form
\begin{equation}
\d s^2 = - \;\exp\left[-2\int_r^\infty g(\tilde r) \; \d\tilde r \right] \; 
\d t^2
+ {\d r^2 \over 1-2\mu(r)\,r^2} +
r^2 \left[ \d\theta^2 +\sin^2\theta\;\d\phi^2\right]
\end{equation}
where $\mu(r) = {4\pi\over3} \bar\rho(r)$ is proportional to the
average density inside radius $r$. In terms of these variables, the
Einstein equations are
\begin{eqnarray}
8\pi \rho &=& G_{\hat t\hat t} = 2 m'(r)/r^2 = 2 [r\,\mu'(r)+3\mu(r)];
\\
8\pi p &=& G_{\hat r\hat r} = 
2 \left\{ 
{g(r) \; \left[1-{2\mu(r) r^2}\right]\over r} - \mu(r)
\right\};
\\
8\pi p &=& G_{\hat \theta\hat \theta} = 
-r[1+rg(r)] \; {\d\mu(r)\over\d r} 
-2 \left\{ [1+r\,g(r)]^2 +r^2\,{\d g(r)\over\d r} \right\}  \mu(r)
+ \left[ {\d g(r)\over\d r} + {g(r)\over r} + g(r)^2 \right].
\end{eqnarray}
The first of these equations integrates to
\begin{equation}
\mu(r) = {1\over r^3} \int_0^r 4\pi \rho(\tilde r) \; \tilde r^2 \;\d\tilde r,
\end{equation}
which justifies the choice of notation $m(r)=\mu(r)\;r^3$. 

\section{General solution and generating function}

By demanding the isotropy condition $G_{\hat r\hat r} =
G_{\hat\theta\hat\theta}$ and algebraically solving for $\d g/\d r$ we
obtain
\begin{equation}
{\d g\over\d r} = 
- g^2 + {1+\mu' r^3\over r(1-2\mu \;r^2)}\; g 
+ {r\;\mu'\over1-2\mu\;r^2}.
\end{equation}
This is a Riccati equation, for which there is no general analytic
solution. If on the other hand we take this \emph{same} equation and
rearrange it algebraically to extract $\d\mu/\d r$ we find
\begin{equation}
{\d\mu\over\d r} = -{2r\,(g^2+g')\over1+r\;g}\; \mu 
+ {(g/r)'+g^2/r\over1+r\;g}.
\end{equation}
But this is a simple first-order linear ODE and hence explicitly
solvable. A symbolic manipulation program such as Maple, or a slightly
tedious hand-computation, easily yields the general solution
\begin{equation}
\mu(r) = \exp\left[-2\int {r\; [g^2(r)+g'(r)]\over1+r g(r)} \;\d r\right]
\; \left\{ C_1 +
\int {-g(r)+ r g'(r) + r g(r)^2\over r^2[1+r g(r)]} \;
\exp\left[+2\int {r\;[g^2(r)+g'(r)]\over1+r g(r)} \; \d r\right]
\right\}.
\label{e:first}
\end{equation}
This statement is equivalent to the algorithm presented by
Lake~\cite{Lake}: Given a prescribed gravity profile $g(r)$ (the
``generating function''), and the knowledge that we are dealing with a
perfect fluid, the mass profile $m(r)=\mu(r)\;r^3$ is deduced in
closed form. The algorithm (\ref{e:first}) is also equivalent to that
presented by Rahman and Visser~\cite{Rahman}, after a change of
coordinates (from isotropic to curvature coordinates) and a change of
variables. A particularly nice feature of the present version of the
algorithm is that the generating function $g(r)$ has a clear physical
interpretation in terms of the gravitational field. Now the above is
by no means the most useful form in which $\mu(r)$ can be presented.
An integration by parts permits us to simplify the appearance of the
integrating factor
\begin{equation}
\exp\left[+2\int {r\;[g^2(r)+g'(r)]\over1+r g(r)} \;\d r\right]
=
[1+r\;g(r)]^2 \; 
\exp\left[ - 2 \int g(r) \; {1-r\;g(r)\over 1+r\;g(r)} \; \d r \right].
\end{equation}
It is now extremely useful to introduce the notation
\begin{equation}
\vartheta(r) = \int g(r) \; {1-r\;g(r)\over 1+r\;g(r)} \; \d r;
\qquad
\hbox{and}
\qquad
\vartheta(r_1;r_2) = 
\int_{r_1}^{r_2} g(r) \; {1-r\;g(r)\over 1+r\;g(r)} \; \d r.
\end{equation}
We warn the reader that we cannot generally assume $r\; g(r) \leq 1$,
and the consequently $\vartheta$ may become negative. For instance, in
the physically reasonable regime $m(r)/r > 1/3$ [that is, $\mu(r) \;
r^2 > 1/3$] and $p\geq0$, it can be shown from the $G_{\hat r\hat r}$
Einstein equation that $r\; g(r) > 1$. All that we can safely say in
general is that as long as local gravity points down we must have
\begin{equation}
- \int g(r)\; \d r < \vartheta < \int g(r) \; \d r.
\end{equation}

With this notation
\begin{equation}
\mu(r) = {\exp[+2\vartheta(r)]\over[1+r\;g(r)]^2} \;
\left\{ 
C_2 + \int [1+r\;g(r)] \; {[-g(r)+ r g'(r) + r g(r)^2]\over r^2}
\; \exp[-2\vartheta(r)] \; \d r 
\right\}.
\end{equation}
A second integration by parts now yields
\begin{equation}
\mu(r) = {\exp[+2\vartheta(r)]\over[1+r\;g(r)]^2} \;
\left\{ 
C_3 
+ {1\over2}\;[1+r\;g(r)]^2 \; {\exp[-2\vartheta(r)]\over r^2}
+ \int {1+r\;g(r)\over r^3}
\; \exp[-2\vartheta(r)] \; \d r
\right\}.
\end{equation}
In this version of the result we have eliminated all the derivatives
of $g(r)$.  A third integration by parts, using $1/r^3 = -{1\over2}
(1/r^2)'$ then leads to
\begin{equation}
\mu(r) = {g(r)\over r} \; {[1+{1\over2}r\;g(r)]\over[1+r\;g(r)]^2} 
+{\exp[+2\vartheta(r)]\over[1+r\;g(r)]^2} \;
\left\{ 
C_4 
+ 2 \int {g(r)^2\over r[1+r\;g(r)]} \; \exp[-2\vartheta(r)] \; \d r
\right\}.
\label{e:final}
\end{equation}
This final version, as we shall soon see, has nice behaviour at the
origin.  Again, we emphasise that this is the explicit, and most
general, solution to the perfect fluid constraint for \emph{arbitrary}
generating function $g(r)$. All perfect fluid spheres, no matter how
derived, must satisfy this equation.

The pressure can now be determined using the $G_{\hat r\hat r}$
Einstein equation so that
\begin{equation}
p(r) = {1\over8\pi[1+r\;g(r)]^2} 
\left[
- g(r)^2 - 2[1+2r\;g(r)] \exp[+2\vartheta(r)]
\left\{ C_4 
+ 2  \int {g(r)^2\over r[1+r\;g(r)]} \; \exp[-2\vartheta(r)] \; \d r
\right\}
\right].
\label{e:pressure}
\end{equation}
This now provides for us the explicit and fully general solution to
the mass profile and pressure profile, given only the gravity profile
and the information that we are dealing with a static spherically
symmetric perfect fluid.  For a consistency check, we can compare
these formulae to the much simpler result for Newtonian stars:
\begin{equation}
\mu(r) = {g(r)\over r},
\qquad
\hbox{and}
\qquad
p(r) = {1\over8\pi} 
\left[ - g(r)^2 + C_5 - 4 \int {g(r)^2\over r} \; \d r \right].
\end{equation}
To complete the analysis we should now impose boundary conditions.
There are three natural locations to work with: (1) the center of the
fluid body, (2) the surface of the fluid body [assuming it has a well
defined surface], and (3) spatial infinity. Perhaps surprisingly, the
simplest results are obtained if we normalize at spatial infinity.

\section{Boundary conditions at spatial infinity}
 
We will now adopt the very mild condition that the total mass of the
fluid sphere is finite, so that $\mu(r)\to0$ as one approaches spatial
infinity. We also assume $p(r)\to 0$ at spatial infinity. Then (from
the $G_{\hat r\hat r}$ equation) we deduce $g(r)\to 0$ at spatial
infinity.  Physically this means that the present discussion is
capable of handling situations with a tenuous atmosphere extending all
the way to infinity, and that the special case where the fluid body
has a sharp surface with $p(r\geq R)=0$ and $m(r\geq R)=M$ is
automatically included.  Then fixing boundary conditions at spatial
infinity, the mass profile (\ref{e:final}) is given by
\begin{equation}
\mu(r) = {g(r)\over r} \; {[1+{1\over2}r\;g(r)]\over[1+r\;g(r)]^2} 
-2{\exp[-2\vartheta(r;\infty)]\over[1+r\;g(r)]^2} \;
\int_r^\infty {g(\tilde r)^2\over \tilde r[1+\tilde r\;g(\tilde r)]} 
\; \exp[+2\vartheta(\tilde r;\infty)] \; \d \tilde r.
\end{equation}
We can simplify this slightly to yield
\begin{equation}
\mu(r) = {g(r)\over r} \; {[1+{1\over2}r\;g(r)]\over[1+r\;g(r)]^2} 
-{2\over[1+r\;g(r)]^2} \;
\int_r^\infty {g(\tilde r)^2\over \tilde r[1+\tilde r\;g(\tilde r)]} 
\; \exp[-2\vartheta(r;\tilde r)] \; \d \tilde r.
\end{equation}
The pressure profile determined from (\ref{e:pressure}) is then
\begin{equation}
p(r) = {1\over8\pi[1+r\;g(r)]^2} 
\left[
- g(r)^2 +4[1+2r\;g(r)] 
\int_r^\infty {g(\tilde r)^2\over \tilde r[1+\tilde r\;g(\tilde r)]} 
\; \exp[-2\vartheta(r;\tilde r)] \; \d \tilde r
\right].
\end{equation}
For the central pressure, $p_c = p(0)$, we find
\begin{eqnarray}
p_c &=&
{1\over2\pi} 
\int_0^\infty {g(r)^2\over r[1+ r\;g(r)]} 
\; \exp[-\vartheta(0;r)] \; \d  r.
\end{eqnarray}
Compare with the equivalent statement for a Newtonian fluid body in
which the pressure profile is
\begin{equation}
p(r) = {1\over8\pi}\left[ 
-g(r)^2 
+ 4 \int_r^\infty {g(\tilde r)^2\over\tilde r} \; \d \tilde r
\right],
\end{equation}
and the central pressure is
\begin{equation}
p_c = {1\over2\pi} 
\int_0^\infty {g(r)^2\over r} \; \d r.
\end{equation}

\section{Boundary conditions at the center of the fluid sphere}
  
If we apply boundary conditions at the center of the sphere then, using
\begin{equation}
g(r) = {m(r) + 4\pi p(r) r^3\over r^2[1-2m(r)/r]}
\end{equation}
and the assumed finiteness of $\rho_c$ and $p_c$, implies
\begin{equation}
g(r) = {4\pi\over3} (\rho_c+3p_c) r + O(r^2).
\end{equation}
The mass and pressure profiles are given by
\begin{equation}
\mu(r) = {g(r)\over r} \; {[1+{1\over2}r\;g(r)]\over[1+r\;g(r)]^2} 
+{\exp[+2\vartheta(0;r)]\over[1+r\;g(r)]^2} \;
\left\{ 
-4\pi p_c + 2
\int_0^r {g(\tilde r)^2\over \tilde r[1+\tilde r\;g(\tilde r)]} 
\; \exp[-2\vartheta(0;\tilde r)] \; \d \tilde r
\right\},
\end{equation}
and
\begin{equation}
p(r) = {1\over8\pi[1+r\;g(r)]^2} 
\left[
- g(r)^2 +[1+2r\;g(r)] \; \exp[+2\vartheta(0;r;)]
\left\{
8\pi \; p_c - 4
\int_0^r {g(\tilde r)^2\over \tilde r[1+\tilde r\;g(\tilde r)]} 
\; \exp[-2\vartheta(0;\tilde r)] \; \d \tilde r
\right\}
\right],
\end{equation}
to be compared with the Newtonian result
\begin{equation}
p(r) = p_c -{1\over8\pi} \left[ 
g(r)^2 
+ 4 \int_0^r {g(\tilde r)^2\over\tilde r} \; \d \tilde r
\right].
\end{equation}

\section{Boundary conditions at the surface of the fluid sphere}
 
If the fluid sphere has a sharp boundary, (say at radius $R$, with the
density and pressure identically zero outside this radius), then it
can be useful to normalize at this surface.  For the pressure profile
we find [in terms of the ``surface gravity'' $g_s =
(M/R^2)/\sqrt{1-2M/R}$\,] that
\begin{equation}
p(r) = {1\over8\pi[1+r\;g(r)]^2} 
\left[
- g(r)^2 +[1+2r\;g(r)] \; \exp[-2\vartheta(r;R)]
\left\{
{g_s^2\over1+2R\;g_s}+
\int_r^R {g(\tilde r)^2\over \tilde r[1+\tilde r\;g(\tilde r)]} 
\; \exp[+2\vartheta(\tilde r;R)] \; \d \tilde r
\right\}
\right].
\end{equation}
There is a similar but uninteresting expression for $\mu(r)$. The
central pressure is now
\begin{equation}
p_c = {1\over8\pi}
\left\{ 
{g_s^2\; \exp[-2\vartheta(0;R)] \over1+2R\;g_s}
+
4 \int_0^R {g(r)^2\over r[1+r\;g(r)]} 
\; \exp[-2\vartheta(0;r)] \; \d r
\right\},
\end{equation}
where we have now reduced the range of integration from $(0,\infty)$
to $(0,R)$ at the price of introducing an extra term depending
explicitly on total mass and radius of the fluid sphere.  This can be
compared with the equivalent Newtonian results
\begin{equation}
p(r) = {1\over8\pi} \left[ g_s^2 -
g(r)^2 
+ 4 \int_r^R {g(\tilde r)^2\over\tilde r} \; \d \tilde r
\right]\;
\qquad
\hbox{with}
\qquad
g_s = {M\over R^2},
\end{equation}
and
\begin{equation}
p_c = {1\over8\pi} 
\left[ 
g_s^2 +  4 \int_0^R {g(r)^2\over r} \; \d r
\right].
\end{equation}

\section{Solution generalization technique}

One nice feature of the present analysis is that it allows one to turn
simple exact solutions into more complicated ones. While the general
algorithm presented above always provides an exact solution, it may
not be an ``elementary'' solution in the sense that the integrations
might not be do-able in terms of either elementary or special
functions.  In such a situation, a simplified algorithm is sometimes
useful.

Suppose that one has found, by some unspecified means, a specific
exact solution for a perfect fluid sphere. Let that exact solution be
given in terms of $m(r)$ [or equivalently $\mu(r)$] and $g(r)$.  Then
for any arbitrary constant $k$,
\begin{equation}
\mu(r) \to \mu(r) + k \;\exp\left[-2\int {r\; [g^2(r)+g'(r)]\over1+r g(r)} \;\d r\right]
\end{equation}
is also an exact solution for a perfect fluid sphere [with the
\emph{same} $g(r)$]. This construction may sometimes ``fail'' in the
sense that the integral is either too trivial [returning you to the
seed solution you started with], or too complicated to perform in
terms of elementary or special functions. However in very many cases
this simple construction is sufficient to understand why certain broad
classes of exact solution exist.

Let us start by rescaling the time variable to remove any redundancies
in the number of free parameters, $n$, appearing in $g_{tt}$. If the
number of free parameters appearing in $g_{rr}$ is not at least $n+1$
then the seed solution you have \emph{must} have a generalization.
For instance the Minkowski solution [a particularly simple fluid
sphere with zero pressure and density] has exactly zero parameters
appearing in both $g_{tt}$ and $g_{rr}$, and so must have a
one-parameter generalization. In this case, performing the integration
leads to the Einstein static universe.  Similarly, the exterior
Schwarzschild solution [another particularly simple fluid sphere with
zero pressure and density] has exactly one free parameter [the mass]
appearing in both $g_{tt}$ and $g_{rr}$, and so must have a
one-parameter generalization. In this case, performing the integration
leads to what is called the Kuch68 II solution in the Delgaty--Lake
classification~\cite{Delgaty}.  A slightly more complex example, using
anti-de Sitter space as a seed, leads to the Tolman IV solution. A
number of additional examples of this phenomena are collected in
table~I.

\begin{table}[ht]
\begin{tabular}{|l|l|}
\hline
Seed & Generalization \\
\hline
Minkowski & Einstein static\\
Schwarzschild exterior& Kuch68 II\\
anti-de Sitter& Tolman IV\\
Tolman V & Kuch2 I\\
Tolman VI& Wyman IIa\\
\hline
Kuch1 Ib & appears new\\
M--W III & appears new\\
K--O III & appears new\\
\hline
\end{tabular}
\caption{Table I: Seed solutions and their generalizations.}
\end{table}

Of course, sometimes explicit exact solutions were first discovered in
their general form, in which case this algorithm provides no extra
information. (This comment applies, for instance to the Wyman IIb
geometry.)  Conversely, sometimes the integral is too complicated to
provide a closed-form solution --- the generalization my be exact but
too complex to write down explicitly. (As for instance when you use
the Schwarzschild--de Sitter [Kottler] geometry as seed. Similarly, by
parameter counting Tolman VII and Tolman VIII must have one-parameter
extensions, but it seems impossible to write then down in closed
form.)

Finally, we point out that there are some cases where this formalism
does lead to apparently new solutions. (We again follow the
Delgaty--Lake classification~\cite{Delgaty}.) For instance, the Kuch1
Ib solution
\begin{equation}
\d s^2 = -(A\;r+B\;r \ln r)^2 \d t^2 + {\d r^2\over 2} + r^2 \; \d\Omega^2
\end{equation}
generalizes to
\begin{equation}
\d s^2 = (A\;r+B\;r \ln r)^2 \d t^2 + 
{  2 (2A + 2B\ln(r) + B) \over 
(2A + 2B\ln(r) + B) - k r^2}\; \d r^2+ r^2 \; \d\Omega^2
\end{equation}
which appears to be new.  Similarly, the M--W III solution, which can be cast
into the form,
\begin{equation}
\d s^2 =  -\left(r-{r^2\over a}\right) \;\d t^2 + {7\d r^2\over 4(1-r^2/a^2)} + r^2 \; \d\Omega^2
\end{equation}
generalizes to
\begin{equation}
\d s^2 =  \left(r-{r^2\over a}\right) \; \d t^2 + {\d r^2\over 1 - 2m(r)/r} + r^2 \; \d\Omega^2
\end{equation}
with
\begin{equation}
m(r) = {4r^2+3a^2\over14a^2} r + k { (r-a) r^{10/3}\over (4r-3a)^{4/3} }
\end{equation}
which also appears to be new.  Also, the K--O III solution can be cast
into the form
\begin{equation}
\d s^2 =  -\left(1+{r^2\over a^2}\right)^2 \; \d t^2 + \d r^2 + r^2 \; \d\Omega^2
\end{equation}
which is spatially flat. It generalizes to
\begin{equation}
\d s^2 =   -\left(1+{r^2\over a^2}\right)^2\; \d t^2 + {\d r^2\over 1 - k r^2 (3r^2+a^2)^{-2/3}} 
+ r^2 \; \d\Omega^2,
\end{equation}
which is contained within the new class of exact solutions briefly
described by Lake~\cite{Lake}.

\section{Discussion}
 
As emphasised in the article by Rahman and Visser~\cite{Rahman}, and
reiterated by Lake~\cite{Lake}, while this type of algorithm
guarantees a perfect fluid body it does not necessarily guarantee a
``physically reasonable'' perfect fluid body.  One physically
reasonable constraint that is easy to enforce in the current
formulation is $g>0$; locally measured gravity should always attract
towards the center of the body.  A second physically reasonable
constraint which is automatically satisfied is that the central
pressure is positive. It is considerably more difficult to enforce
$m(r)\geq 0$, $\rho(r)\geq 0$, and $p(r)\geq 0$.  Checking these
physically motivated constraints amounts to mathematically
investigating a set of integral inequalities, and seems to require a
case by case investigation depending on the assumed gravity profile
$g(r)$.  One should not however lose track of the significance of what
has been accomplished:
\begin{itemize}
\item We have derived the exact and fully general solution to the
  pressure isotropy condition in terms of variables that have a direct
  physical meaning, the gravity profile $g(r)$ and mass profile
  $m(r)$.
\item We have also derived an exact and fully general formula for the
  pressure profile $p(r)$ of a perfect fluid sphere that depends only
  on the gravity profile $g(r)$.
\item In particular we have an exact and fully general expression for
  the central pressure of a fluid sphere, again determined directly in
  terms of the gravity profile $g(r)$.
\item The algorithm provides a natural framework for understanding the
  reason for the existence of certain broad classes of exact solution,
  and in some cases leads to new exact solutions.
\end{itemize}
Because this algorithmic approach works directly in terms of
physically meaningful quantities, with a physically meaningful
``generating function'' in the form of the gravity profile $g(r)$, the
interpretation of the results is somewhat clearer than in the
algorithms presented in the Rahman--Visser~\cite{Rahman} and
Lake~\cite{Lake} articles. We expect that this version of the
algorithm for generating perfect fluid spheres will lead to additional
useful ``exact solutions''. In particular, the new class of exact
solutions briefly described in~\cite{Lake} has a very natural
representation in terms of this algorithm.

\acknowledgments

This Research was supported by the Marsden Fund administered by the
Royal Society of New Zealand.



\end{document}